\documentclass[cits]{PoS}

\usepackage{multirow}
\usepackage{url}
\usepackage{amsmath, amssymb}
\usepackage{epsf,epsfig,graphics}
\usepackage{subfig}

\def\LALPHA{\hbox{\epsfxsize=2.0 true cm
    \epsfbox{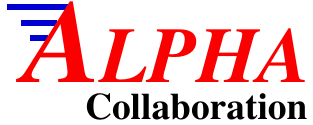}}
}
\newcommand{\onecol}[2]{
  \begin{minipage}[t]{#1}{#2\vfill} \end{minipage}
}

\title{The $B^*B\pi$ Coupling in the Static Limit}

\ShortTitle{The $B^*B\pi$ Coupling}

\author{\LALPHA \hfill
  \onecol{4.0cm}{\vspace{-1.5cm}\it DESY 10-194 \\ SFB/CPP-10-103
    \vspace{-2.1cm}
  }}

\author{J. Bulava, \speaker{M.A. Donnellan}, Rainer Sommer\\
  NIC, DESY, Platanenallee 6, 15738 Zeuthen, Germany\\
  E-mail: \email{michael.donnellan@desy.de}}

\abstract{We study an accurate method for the lattice calculation of
  the $B^*B\pi$-coupling in the static limit, paying particular
  attention to excited state contamination. As this coupling is a
  parameter of the heavy meson chiral Lagrangian, it is useful for
  constraining the chiral behaviour of various observables in
  $B$-physics. We present a precise study of the continuum limit in
  the quenched approximation and preliminary results with 2 flavours
  of improved Wilson quarks (using CLS lattices) for pion masses down
  to around 250 MeV. With dynamical quarks both the lattice spacing
  and the light quark mass dependences are found to be very weak. We
  can quote $g$ = 0.51(2) for the continuum value in the chiral limit,
  where the error will be reduced when a more complete analysis has
  been performed.}

\FullConference{The XXVIII International Symposium on Lattice Field Theory\\
  June 14-19,2010\\
  Villasimius, Sardinia Italy}

\begin{document}

\section{Introduction}

The $B^*B\pi$-coupling is defined via the matrix element associated
with the strong decay $B^*\rightarrow\;B\pi$:
\begin{equation}
  \langle B^0(p)\pi^+(q)|B^{*+}(p')\rangle \equiv
  -{g_{B^*B\pi}}q_\mu\eta^\mu(p')(2\pi)^4\delta(p'-p-q),
\end{equation}
where $\eta^\mu$ is the polarization vector of the $B^*$ meson. With
the physical quark masses $m_{B^*} < m_B + m_\pi$, so this decay is
kinematically forbidden and direct experimental measurement of the
coupling is not possible. The equivalent charmed meson decay, $D^*
\rightarrow D \pi$, is however the dominant decay mode of the $D^*$
meson, and experimental measurements of the corresponding coupling are
available: $g_{c} = 0.61(7)$ \cite{Anastassov:2001cw}.

Much of the interest in this coupling is due to the role it plays in
heavy-light meson chiral perturbation theory (HM$\chi$PT)
\cite{Casalbuoni:1996pg}. At leading order in the heavy quark and
chiral expansions, there is in addition to the pion decay constant
$f_\pi$ a single low-energy constant $g$ which has a simple relation
to the $B^*B\pi$-coupling:
\begin{equation}
  g = \lim_{m_b\rightarrow\infty, m_d\rightarrow 0} \frac{g_{B^*B\pi}}{2\sqrt{m_B m_{B^*}}}f_\pi.
\end{equation}
Knowledge of this coupling is thus helpful for constraining the chiral
behaviour of various $B$-physics observables. Since in the ALPHA
Collaboration HQET programme we will calculate several such
quantities, e.g. $f_B$ and the $B \rightarrow \pi l \nu$ form factor,
we begin with a precision determination of $g$.

In lattice calculations of the $B^*B\pi$
coupling\cite{deDivitiis:1998kj,Ohki:2008py,Becirevic:2009yb}, the
issue of multihadronic $|B\pi\rangle$ states can be avoided entirely
by using a combination of LSZ reduction of the pion and pion dominance
in order to relate the coupling to the form factor of the axial
current between $B$ and $B^*$ states. This reduction of the pion is
analogous to the Golberger-Treiman relation for the
nucleon-nucleon-pion coupling. The form factor can then be written:
\begin{equation}
  \langle B^0(p)|A_\mu|B^{*+}(p+q)\rangle = \eta_\mu F_1(q^2) + (\eta \cdot q)(2p+q)_\mu F_2(q^2) + (\eta\cdot q)q_\mu F_3(q^2)
\end{equation}
In the static and chiral limits, in which the leading-order HM$\chi$
coupling is obtained, only the zero-momentum form factor $F_1(0)$ must
be calculated.

\section{Method}

The lattice determination of a form factor requires the calculation of
three-point functions. Particular attention must therefore be paid to
the problems of statistical noise (which grows exponentially with the
Euclidean time separations in heavy-light correlation functions) and
excited state contamination (which is exponentially suppressed with
Euclidean time separation).

In a generic lattice three point function:
\begin{equation}
  C_3(t,t';\alpha,\beta) = \langle {\mathcal O}_\alpha(t) {\mathcal O}(t') {\mathcal O}^\dagger_\beta (0)\rangle,
\end{equation}
where $\beta$ and $\alpha$ label the quantum numbers of the
interpolating fields at the source and sink respectively, there are 2
time separations $t'$ and $t-t'$ which ought both to be taken as large
as possible if excited state contamination is to be minimized.  For
example, in the case of a matrix element between states with equal
energies, putting the insertion point midway between the source and
the sink we have $C_3(t,t/2;\alpha,\beta)$ which has corrections
${\mathcal O}(e^{-(t/2)\Delta E})$ with the energy gap $\Delta E =
E_2^{(\alpha)}-E_1^{(\alpha)} = E_2^{(\beta)}-E_1^{(\beta)}$.

In this work we investigate a method in which we calculate instead,
for a given source-sink separation, the sum of the correlation
functions with the operator inserted on each timeslice.
\cite{Maiani:1987by,Gusken:1988yi}. This helps to reduce
excited state contamination. Defining the summed correlator by:
\begin{equation}
  D(t;\alpha,\beta) \equiv a\sum_{t'} C_3(t,t';\alpha,\beta),
\end{equation}
it can be shown that the form of the corrections in the effective
matrix element is:
\begin{equation}
  R(t) \equiv \partial_t \frac{D(t;\alpha,\beta)}{\sqrt{C_2(t;\beta)C_2(t;\alpha)}} = \langle\beta|{\mathcal O}|\alpha\rangle + {\mathcal
  O}(te^{-t\Delta E}),
  \label{eqn:Rdef}
\end{equation}
where the two-point function is given by $C_2(t;\alpha) \equiv
\langle{\mathcal O}_\alpha(t){\mathcal O}^\dagger_\alpha(0)\rangle$.

In the calculation of the bare matrix element we use both the HYP1 and
HYP2 discretizations of the static action in order to mitigate the
exponential decay of signal-to-noise in Euclidean time
\cite{DellaMorte:2003mn,DellaMorte:2005yc} and also to get an
additional handle on discretization effects. We also use noisy
estimation of the all-to-all light quark
propagator\cite{Sommer:1994gg}, with U(1) noise and full
time-dilution. Since we use a sequential propagator for the light
quark, we obtain easily the insertion summation over all timeslices
$t'$ but must invert again for each operator used (in this case we
perform the insertion for each spatial component of the axial
current).

Additionally, we use a variational basis of 8 interpolating operators
for the $B$ and $B^*$ mesons, consisting of Gaussian smeared operators
with an APE-smeared gauge covariant Laplacian $\Delta$:
\begin{equation}
  \psi_l^{(k)}(x) = \left(1 + \kappa_G a^2 \Delta\right)^{R_k}\psi_l(x)
\end{equation}
with widths
\begin{equation}
  r_{phys,k} \approx 2\sqrt{\kappa_G R_k}a
\end{equation}
varied over the range 0 - 0.7 fm. In order to extract the desired
matrix element from the matrices of correlation functions obtained in
this way, we use a Generalized Eigenvalue Problem (GEVP) analysis in
order to construct optimized interpolating operators\cite{Michael:1982gb}.

\begin{table}[t]
  \centering
  \begin{tabular}{ c | c | c}
    Ensemble parameters&$N_{conf}$&$N_{noise}$\\\hline
    $16^3\times 32$, $\beta=6.0219$, $a\approx 0.10$ fm& 100 & 200 \\
    $24^3\times 48$, $\beta=6.2885$, $a\approx 0.08$ fm& 100 & 48 \\
    $32^3\times 64$, $\beta=6.4956$, $a\approx 0.05$ fm& 100 & 32\\
  \end{tabular}
  \caption{Details of the quenched lattices and
    measurements. $N_{conf}$ gives the number of configurations and $N_{noise}$ the number of noise sources per
    timeslice.}
  \label{tab:q}
\end{table}

\section{Quenched Tests and Results}
We have performed a test of our method using the lattices and valence
strange quark parameter $\kappa_s$ of the ALPHA Collaboration quenched
HQET programme\cite{Blossier:2010jk,Blossier:2010vz,Blossier:2010mk}.
The Wilson gauge action is used and there are 3 lattice spacings with
a fixed physical volume of $L\approx 1.5$ fm, and temporal extent
$T=2L$.  The details of the lattices and of the measurements performed
on them are given in Table~\ref{tab:q}.

In each case, it was possible to determine the value of the bare
matrix element with sub-percent precision. The renormalized
axial current is given however by:
\begin{equation}
  (A_k)_R =  Z_A(g_0^2)(1+b_A(g_0^2)am_q)(A_k + ac_A(g_0^2)\partial_k P),
\end{equation}
where $P$ represents the pseudoscalar density. At zero momentum
transfer $\vec{q}=0$, the $c_A$ term does not contribute. We use the values of
$\kappa_c$ and $Z_A$ given in Ref.~\cite{Luscher:1996jn} and the
one-loop perturbative estimate of $b_A$\cite{Sommer:2006sj}. The
overall uncertainty in the calculation is therefore dominated by the
$1\%$ error on $Z_A$.

\begin{figure}[t]
  \centering
  \includegraphics[angle=90,width=12cm]{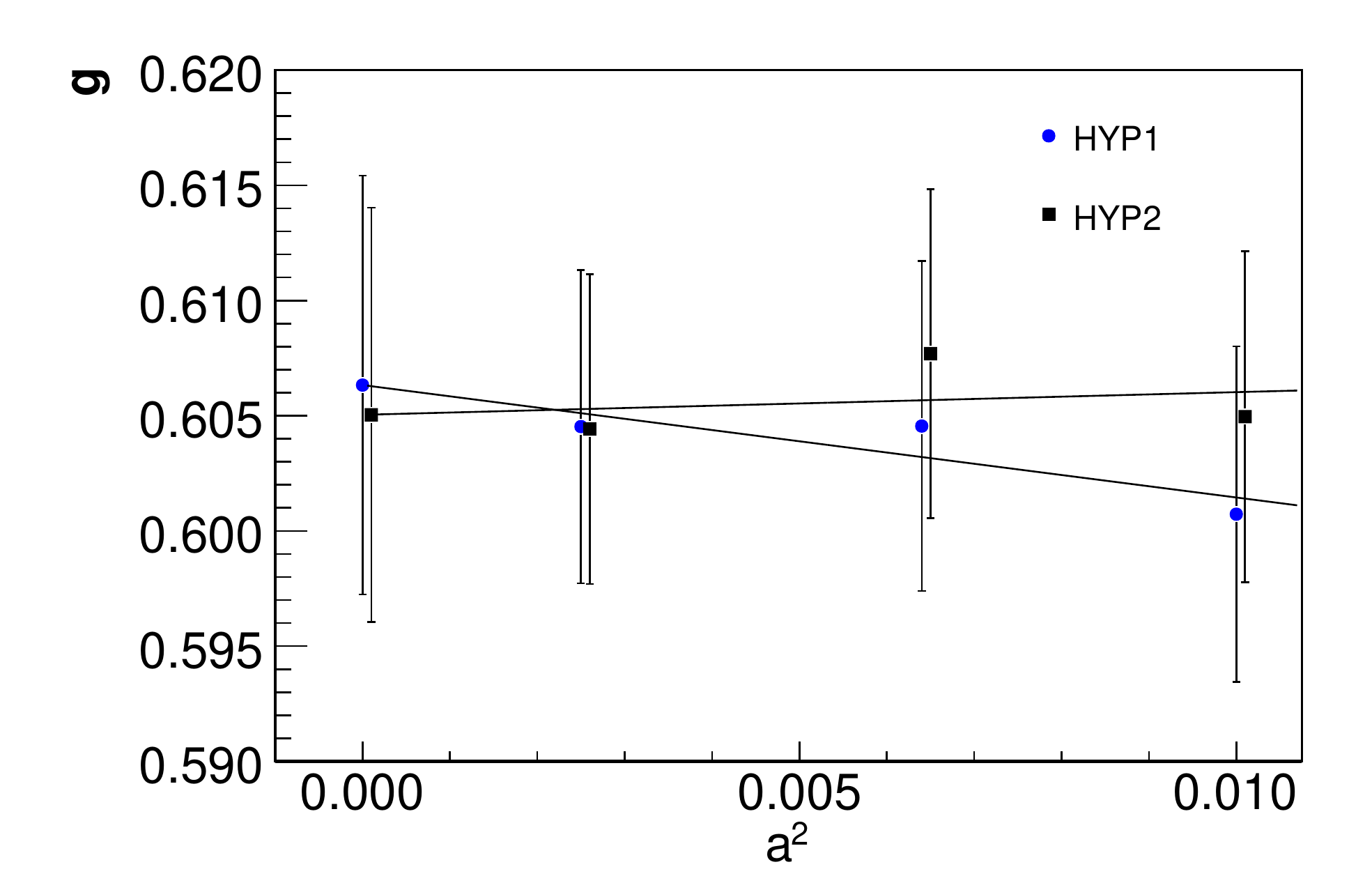}
  \caption{Quenched continuum extrapolation. A horizontal offset has
    been added to the HYP2 points for clarity.}
  \label{fig:q-cont-lim}
\end{figure}

As can be seen in the continuum extrapolation shown in
Figure~\ref{fig:q-cont-lim}, there is no discernible lattice spacing
dependence at this precision. Fitting linearly in $a^2$, we obtain the
continuum limit results:

\begin{equation}
  g^{\mathrm{}}_{\mathrm{HYP1}}=\;0.606(9),\quad\quad g^{\mathrm{}}_{\mathrm{HYP2}}=\;0.605(9).
\end{equation}

\section{$N_f=2$ Results}
For the $N_f=2$ calculation we use gauge configurations generated and
shared within the CLS (``Coordinated Lattice Simulations'') community
effort\footnote{\url{https://twiki.cern.ch/twiki/bin/view/CLS/WebHome}}.
The action is non-perturbatively $O(a)-$improved $N_f = 2$ Wilson QCD,
and the algorithm used is deflation-accelerated DD-HMC
\cite{Luscher:2005rx,Luscher:2007es}. Details of the lattices used and
of the measurements performed so far are given in Table~\ref{tab:nf2}.

\begin{table}[th]
\centering
  \begin{tabular}{c | c | c | c | c | c | c}
    &Name& $L$ (fm) &$m_\pi$ (MeV)&$m_\pi L$ &$N_{conf}$&$N_{noise}$\\\hline
    \multirow{2}{64pt}{$\beta=5.2$\\$a = 0.08$ fm }& A1 & 2.6 & 700 & 9.0 & 288 & 8 \\
    & A4 & 2.6 & 360 & 4.8 & 370 & 8 \\
    \hline
    \multirow{4}{64pt}{$\beta=5.3$\\$a = 0.07$ fm }& E4 & 2.3 & 540 & 6.2 & 157 & 16 \\
    & E5 & 2.3 & 410 & 4.7 & 2000 & 4 \\
    & F6 & 3.4 & 280 & 5.0 & 339 & 4 \\
    & F7 & 3.4 & 250 & 4.2 & 301 & 4 \\
      \hline
      \multirow{2}{64pt}{$\beta=5.5$\\$a = 0.05$ fm }& M5 & 1.6 & 430 & 3.0 & 258 & 12 \\
      & N5 & 2.4 & 430 & 5.3 & 477 & 2 \\
    \end{tabular}
    \caption{Details of the $N_f=2$ lattices and measurements. The scale setting is based on $f_K$ and is preliminary.}
  \label{tab:nf2}
\end{table}

\begin{figure}[h!]
  \centering
  \subfloat[E5 ensemble.]{\includegraphics[angle=90,width=15cm]{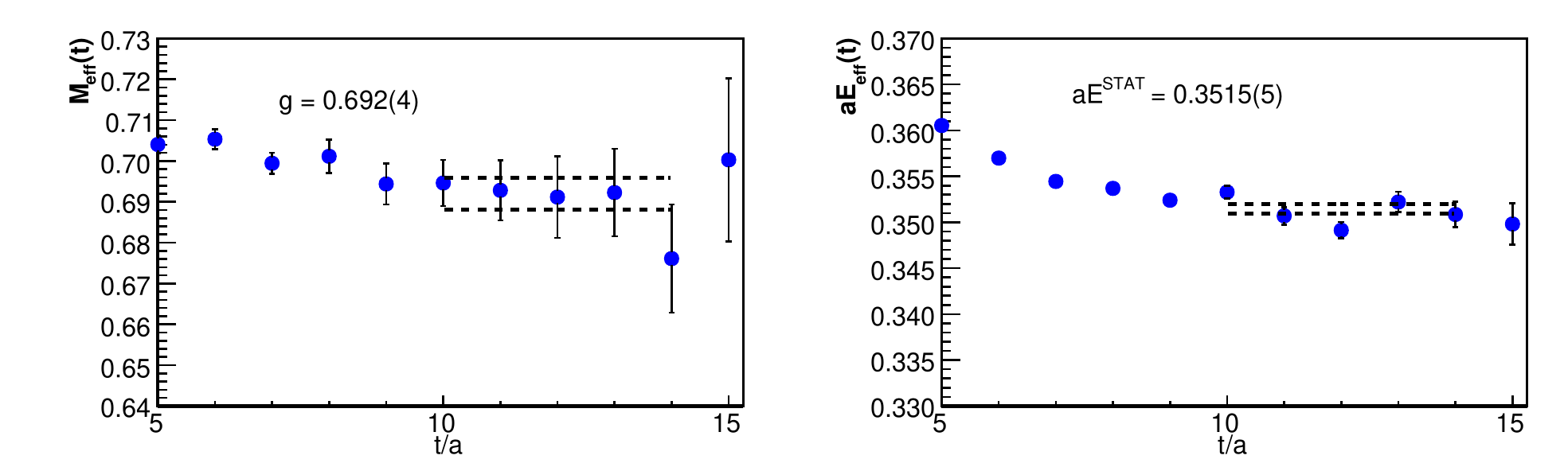}}

  \subfloat[F6 ensemble.]{\includegraphics[angle=90,width=15cm]{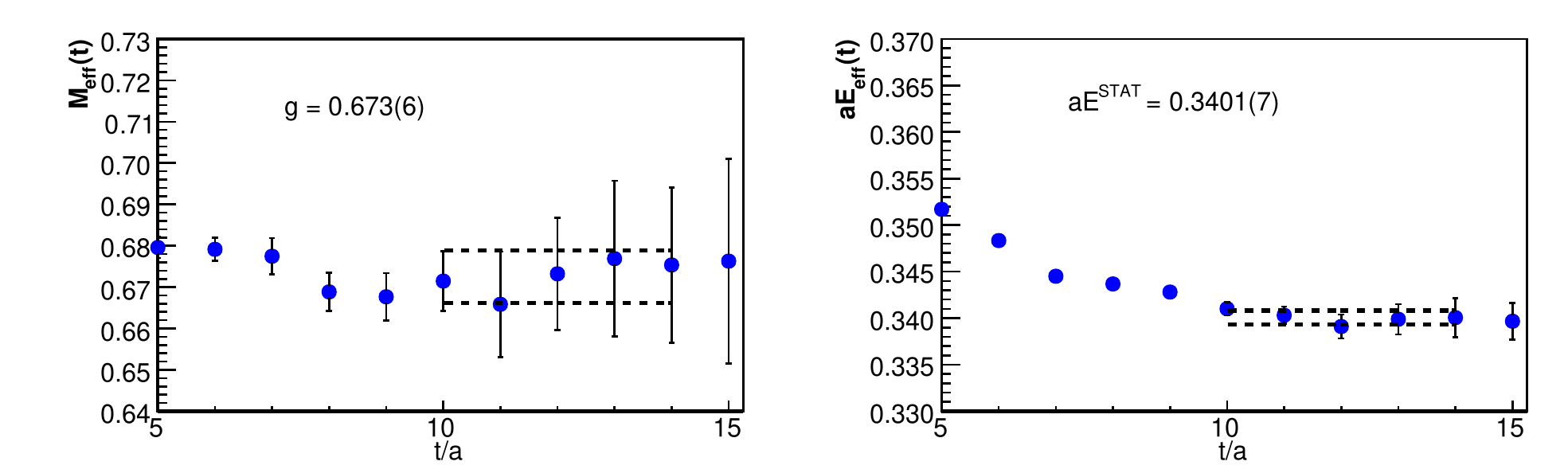}}
  \caption{Example plateaus and bare results. The left panels show the
    ratio $R(t)$ defined in Eqn.~\protect\ref{eqn:Rdef} as well as
    fits to the bare matrix element. The right panels show the
    corresponding two-point function effective mass (the static
    energy). All plateau ranges are chosen by examination of the
    two-point function effective mass. The GEVP initial timeslice
    $t_0$ is chosen to be at 0.3 fm.}
  \label{fig:plateaus}
\end{figure}

Example plateaus for the E5 and F6 ensembles with the HYP2 static
action are shown in Fig.~\ref{fig:plateaus}. For now, we do not
explicitly estimate the autocorrelation time but take 50 jackknife
bins in all cases, which gives a reasonable estimate of the error
judging by the investigation in Ref.~\cite{Schaefer:2010hu}. As in the
quenched case, we are able to determine the bare matrix elements with
sub-percent precision in most cases, and the overall uncertainty is
dominated by that on $Z_A$\cite{DellaMorte:2008xb}. We use the value
of $\kappa_c$ given in Ref.~\cite{DellaMorte:2007sb} and for $b_A$ we
again take the one-loop perturbative estimate from
Ref.~\cite{Sommer:2006sj}.

\begin{figure}[h]
  \centering
  \includegraphics[width=12cm]{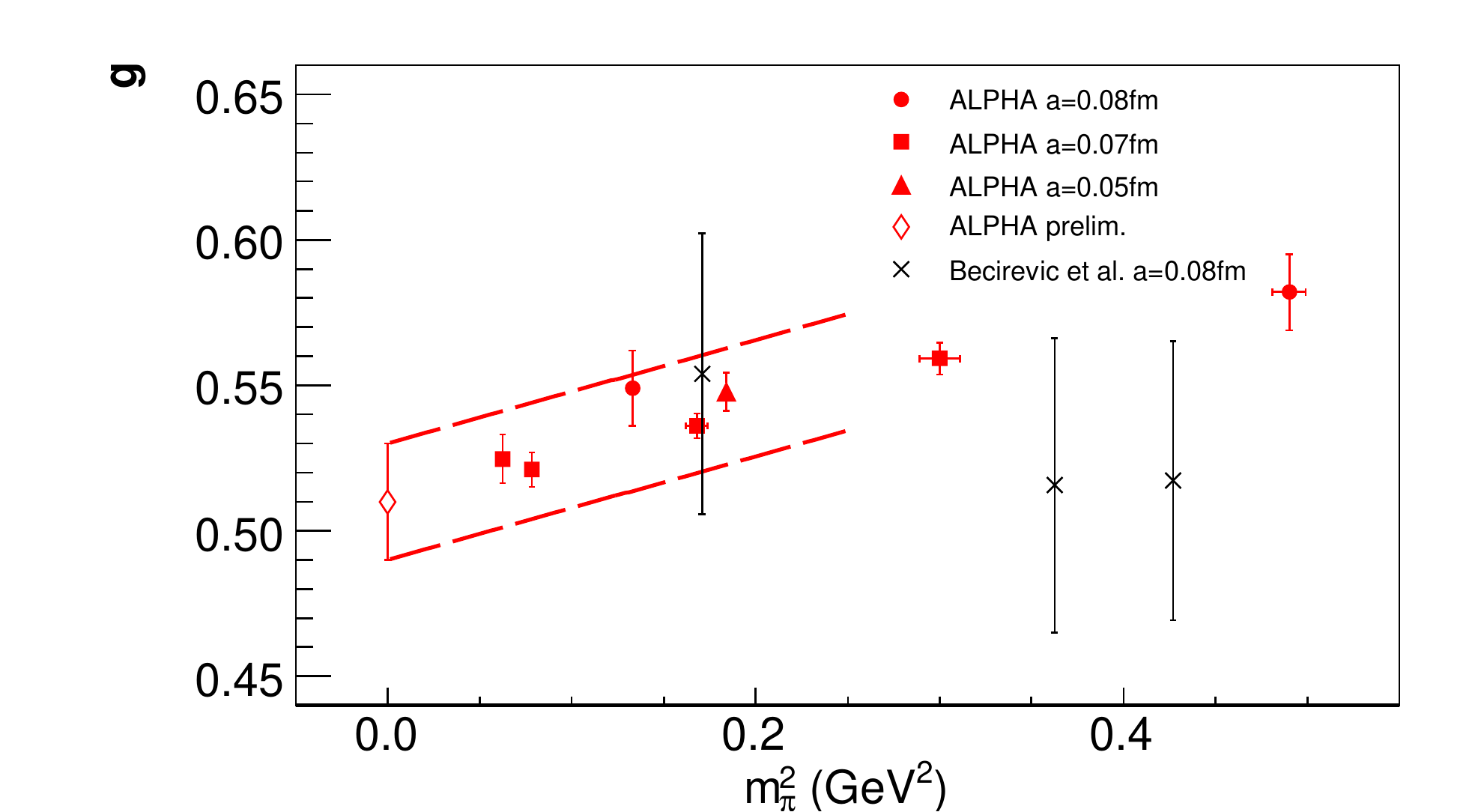}
  \caption{The current status of our renormalized $N_f=2$ results,
    together with those of Ref.~\cite{Becirevic:2009yb} for
    comparison.}
  \label{fig:status}
\end{figure}

In Fig.~\ref{fig:status}, we summarize our renormalized results for
the various ensembles and compare them to those of the recent
publication Ref.~\cite{Becirevic:2009yb}. The results shown use the
HYP2 static action but those for HYP1 look essentially the same. Our
results at $\beta=5.2$ have large errors relative to those at finer
lattice spacings due to the $\approx 2\%$ uncertainty in $Z_A$ at that
value of the bare coupling (compared to $\approx 0.6\%$ at $\beta
=5.3$, for example). At the time of the conference, we had a single
result at $\beta=5.5$ corresponding to the small volume M5 ensemble
which suffers from finite volume effects, and which appeared
significantly lower than those at the coarser lattice spacings. This
point was $10\%$ smaller than the results for the N5 ensemble that are
now available and is therefore excluded from Fig.~\ref{fig:status}.

\section{Summary}
We have performed a lattice study of the $B^*B\pi$ coupling in the
static limit, from which the leading-order HM$\chi$ coupling can be
obtained. Our method combines a summed-insertion three-point function
with a GEVP analysis in order to minimize excited state contamination.
In addition, more standard techniques such as the HYP2 static action
and noisy all-to-all light-quark propagators with timeslice sources
are used, in order to deal efficiently with statistical noise.

We have studied the continuum limit of this quantity in the quenched
approximation with $<2\%$ precision, and see no evidence of lattice
spacing dependence over the range $0.05\;\mathrm{fm} < a <
0.10\;\mathrm{fm}$. In the $N_f=2$ theory we obtain a comparable
precision, surpassing by far that of previous lattice calculations of
this quantity, and have made substantial progress towards controlling
all sources of systematic uncertainty. It is already safe to quote, in
the chiral and continuum limits:
\begin{equation}
g = 0.51(2).
\end{equation}

\section*{Acknowledgements}
We thank Hubert Simma and Fabio Bernardoni for useful discussions.
This work is supported by the Deutsche Forschungsgemeinschaft in the
SFB/TR 09 and by the European community through EU Contract No.
MRTN-CT-2006-035482, ``FLAVIAnet''. We are grateful to NIC and to the
Zuse Institute Berlin for allocating computing resources to this
project. Some of the correlation function measurements were performed
on the PAX cluster at DESY, Zeuthen.

%\begin{thebibliography}{99}
%\bibitem{...} ....
%\end{thebibliography}
\bibliographystyle{h-physrev3}
\bibliography{tau}

\begin{thebibliography}{10}

\bibitem{Anastassov:2001cw}
CLEO, A.~Anastassov {\em et~al.},
\newblock Phys. Rev. {\bf D65}, 032003 (2002), hep-ex/0108043.
%%CITATION = HEP-EX/0108043;%%

\bibitem{Casalbuoni:1996pg}
R.~Casalbuoni {\em et~al.},
\newblock Phys. Rept. {\bf 281}, 145 (1997), hep-ph/9605342.
%%CITATION = HEP-PH/9605342;%%

\bibitem{deDivitiis:1998kj}
UKQCD, G.~M. de~Divitiis {\em et~al.},
\newblock JHEP {\bf 10}, 010 (1998), hep-lat/9807032.
%%CITATION = HEP-LAT/9807032;%%

\bibitem{Ohki:2008py}
H.~Ohki, H.~Matsufuru, and T.~Onogi,
\newblock Phys. Rev. {\bf D77}, 094509 (2008), 0802.1563.
%%CITATION = 0802.1563;%%

\bibitem{Becirevic:2009yb}
D.~Becirevic, B.~Blossier, E.~Chang, and B.~Haas,
\newblock Phys. Lett. {\bf B679}, 231 (2009), 0905.3355.
%%CITATION = 0905.3355;%%

\bibitem{Maiani:1987by}
L.~Maiani, G.~Martinelli, M.~L. Paciello, and B.~Taglienti,
\newblock Nucl. Phys. {\bf B293}, 420 (1987).
%%CITATION = NUPHA,B293,420;%%

\bibitem{Gusken:1988yi}
S.~G{\"u}sken, K.~Schilling, R.~Sommer, K.~H. Mutter, and A.~Patel,
\newblock Phys. Lett. {\bf B212}, 216 (1988).
%%CITATION = PHLTA,B212,216;%%

\bibitem{DellaMorte:2003mn}
ALPHA, M.~Della~Morte {\em et~al.},
\newblock Phys. Lett. {\bf B581}, 93 (2004), hep-lat/0307021.
%%CITATION = HEP-LAT/0307021;%%

\bibitem{DellaMorte:2005yc}
M.~Della~Morte, A.~Shindler, and R.~Sommer,
\newblock JHEP {\bf 08}, 051 (2005), hep-lat/0506008.
%%CITATION = HEP-LAT/0506008;%%

\bibitem{Sommer:1994gg}
R.~Sommer,
\newblock Nucl. Phys. Proc. Suppl. {\bf 42}, 186 (1995), hep-lat/9411024.
%%CITATION = HEP-LAT/9411024;%%

\bibitem{Michael:1982gb}
C.~Michael and I.~Teasdale,
\newblock Nucl. Phys. {\bf B215}, 433 (1983).
%%CITATION = NUPHA,B215,433;%%

\bibitem{Blossier:2010jk}
B.~Blossier, M.~della Morte, N.~Garron, and R.~Sommer,
\newblock JHEP {\bf 06}, 002 (2010), 1001.4783.
%%CITATION = 1001.4783;%%

\bibitem{Blossier:2010vz}
Alpha, B.~Blossier {\em et~al.},
\newblock JHEP {\bf 05}, 074 (2010), 1004.2661.
%%CITATION = 1004.2661;%%

\bibitem{Blossier:2010mk}
ALPHA, B.~Blossier {\em et~al.},
\newblock (2010), 1006.5816.
%%CITATION = 1006.5816;%%

\bibitem{Luscher:1996jn}
M.~L{\"u}scher, S.~Sint, R.~Sommer, and H.~Wittig,
\newblock Nucl. Phys. {\bf B491}, 344 (1997), hep-lat/9611015.
%%CITATION = HEP-LAT/9611015;%%

\bibitem{Sommer:2006sj}
R.~Sommer,
\newblock (2006), hep-lat/0611020.
%%CITATION = HEP-LAT/0611020;%%

\bibitem{Luscher:2005rx}
M.~L{\"u}scher,
\newblock Comput. Phys. Commun. {\bf 165}, 199 (2005), hep-lat/0409106.
%%CITATION = HEP-LAT/0409106;%%

\bibitem{Luscher:2007es}
M.~L{\"u}scher,
\newblock JHEP {\bf 12}, 011 (2007), 0710.5417.
%%CITATION = 0710.5417;%%

\bibitem{Schaefer:2010hu}
S.~Schaefer, R.~Sommer, and F.~Virotta,
\newblock (2010), 1009.5228.
%%CITATION = 1009.5228;%%

\bibitem{DellaMorte:2008xb}
M.~Della~Morte, R.~Sommer, and S.~Takeda,
\newblock Phys. Lett. {\bf B672}, 407 (2009), 0807.1120.
%%CITATION = 0807.1120;%%

\bibitem{DellaMorte:2007sb}
M.~Della~Morte {\em et~al.},
\newblock PoS {\bf LAT2007}, 255 (2007), 0710.1263.
%%CITATION = 0710.1263;%%

\end{thebibliography}
\end{document}